# C$_{60}$-BASED COMPOSITES IN VIEW OF TOPOCHEMICAL REACTIONS. III. C$_{60}$ + GRAPHENE NANOBUDS

E.F.Sheka and L.Kh.Shaymardanova

Peoples' Friendship University of Russia
sheka@icp.ac.ru

**ABSTRACT**: The current paper presents the third part of the study devoted to composites formed by fullerene C$_{60}$ and two graphene nanosheets ([C$_{60}$+(5,5)] and ([C$_{60}$+(9, 8)] graphene nanobuds). The formation of composites is considered from the basic points related to the atomic chemical reactivity of the fullerene molecule and nanographene. The barrier that governs the composites formation is determined in terms of the coupling energy $E_{cpl}^{tot}$ and is expanded over two contributions that present the total energy of deformation of the composites' components $E_{def}^{tot}$ and the energy of covalent coupling $E_{cov}^{tot}$.

In view of these energetic parameters and in contrast to expectations, seemingly identical reactions that are responsible for the formation of intermolecular contacts in (C$_{60}$)$_2$ dimer, [C$_{60}$+(4,4)] carbon nanobuds, and [C$_{60}$+(5,5)] and [C$_{60}$+(9,8)] graphene nanobuds result in different final products. The peculiarity is suggested to result from a topochemical character of the covalent coupling between two members of the $sp^2$ nanocarbons' family. The computations were performed by using the AM1 semiempirical version of unrestricted broken symmetry Hartree-Fock approach.

## 1. Introduction

In contrast to fullerene oligomers and carbon nanobuds, no indication of the existence of chemically bound fullerene-graphene compositions has been so far obtained. It is difficult to believe that no attempts to produce these intriguing composites have been undertaken. Serious reasons appear to make not possible to achieve the goal. Computations performed by Wu and Zeng [1] were first to lift the veil above the feature. They have shown that the reaction of covalent addition of C$_{60}$ to graphene basal plane is endothermic and requires a considerable amount of energy in contrast to C$_{60}$ dimer and [C$_{60}$+(4,4)] CNBs discussed in [2, 3]. However, the mentioned computations have been carried out in a standard configuration of the spin-restricted PBC DFT approach in spite of the fact pointed by the authors themselves that local and semilocal functionals in DFT generally give poor description of weak interaction. Similarly insufficient is response of the technique to the interaction weakening, which takes place between odd electrons of graphene that is why test calculations of the authors within a spin-unrestricted method could not show any difference from the spin-restricted ones due to overpressing the configurational part of the functionals. Performed in the current study, similar test within the framework of the Hartree-Fock approach results in 23% (or 641.6 kcal/mol by absolute value) lowering of the total energy of the (9,9) nanographene (the nomenclature follows the suggested in [4]), which was chosen as supercell in the mentioned PCB DFT computations [1], when going from RHF to UHF approach.

Describing possibility to arrange periodic graphene nanobuds (GNBs) [1], the authors have concentrated their attention on the basal plane of (9,9) nanographene leaving aside the sheet edges as well as supposing homogeneous chemical activity of carbon atoms through over the sheet. It is actually not the case since the chemical activity distribution over graphene atoms is highly inhomogeneous and its edges can be extremely reactive [5-7] thus playing the main role in providing covalent coupling of any addend including carbon nanotubes [5] and $C_{60}$ [8]. This is particularly important when elaborating technology of producing graphene-based nanocarbon composites in solution. To eliminate features related to definite peculiarities in the interaction of graphene with fullerene $C_{60}$, we have performed a computational synthesis of $[C_{60}+(5,5)]$ GNBs varying the place of contact of the molecule with (5,5) nanographene.

Presented below concerns the formation of *AB* product in terms of the scheme shown in Fig.1 in Part 1[2]. The product is formed when the starting intermolecular C-C distances are less than 2.0Å. At longer distances, the pair of $C_{60}$ and nanographene forms a classical charge transfer complex where graphene's atoms contribute into the HOMO while $C_{60}$'s atoms govern the LUMO, which causes the charge transfer from graphene to fullerene under photoexcitation.

## 2. Computational synthesis of graphene nanobuds

Similarly to fullerenes and CNTs, the length of C-C bonds in graphene noticeably exceeds the critical value at which a complete covalent bonding of the relevant odd electrons is terminated [9] so that odd electrons of graphene become effectively unpaired thus providing quite valuable molecular chemical susceptibility $N_D$ of the graphene molecule and atomic chemical susceptibility (ACS) $N_{DA}$ related to each atom. Distributed over the graphene atoms, $N_{DA}$ maps the chemical activity of the molecule atoms. Figure 1 presents the ACS distributions over atoms of (5,5) nanographene under conditions when the sheet edges are either non-terminated (empty) (a) or hydrogen-terminated by one (single-H) (b) and two (double-H) (c) atoms per one carbon. The color pictures present 'chemical portraits' of the three molecules while plotting in Fig. 1d discloses the ACS distributions by absolute values. The total number of unpaired electrons $N_D$ constitutes $31.04e$, $14.57e$, and $12.23e$ in the three cases, respectively. As seen in the figure, the portraits diverge considerably exhibiting the difference in both molecular and atomic chemical activity making the three molecules absolutely different with respect to the same chemical reactions. Non-terminated sheet is the most reactive. Then follow single-H- and double-H terminated ones, the latter is the least active with respect to the total molecular susceptibility. But as seen in Fig.1c (right) and than confirmed in Fig.1d, the reactivity of atoms with maximum ACS values in the latter case is higher than for single-H terminated sample. The feature highlights very important question to be answered before any additive reaction, aimed at producing a wished composite, starts: what is the atomic composition in the edge zone of nanographene samples under study? In the case of $C_{60}$, its addition to the graphene sheet will obviously occur quite differently depending on particular-edge sample and the place of contact to the latter. Let us consider possible situations expected for the above cases basing on the relevant ACS maps.

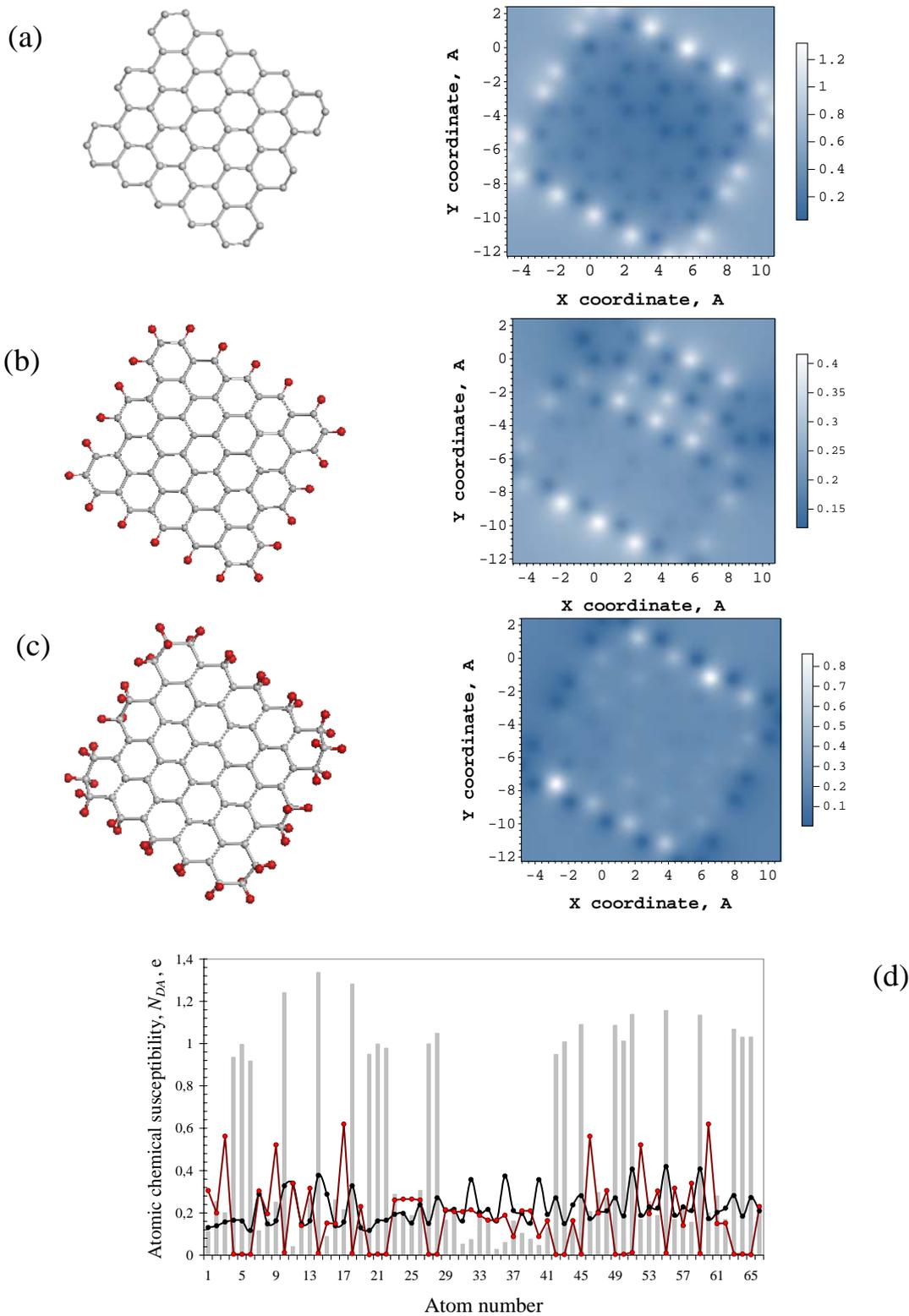

**Figure 1**. Equilibrium structures (left) and chemical portraits (right) of (5,5) nanographene with empty (a) and hydrogen-terminated edge atoms by one (b) and two (c) terminators per carbon atom. Vertical scales determine the $N_{DA}$ values amplitude.
Distribution of the chemical susceptibility over nanographene atoms (d): light gray histogram (case *a*), black curve with dots (case *b*) and dark red curve with dots (case *c*).

## 2.1. Deposition of $C_{60}$ on nanographene with non-terminated edges

According to the ACS map in Fig.1a, atoms of zigzag empty edges are the first targets for the fullerene addition. Approaching to the edge, fullerene molecule will orient itself in such a way to provide the closeness between its most reactive atoms with two zigzag carbon atoms as shown for the relevant GNB 1 in Fig.2. Equilibrium structure of the formed GNB is shown next to the start configuration alongside with the ACS maps related to GNB 1. Energetic parameters are presented in Table 1. We shall refer so far to the total coupling energy $E_{cpl}^{tot}$ only leaving the discussion of other quantities to the next section. The coupling energy is determined as

$$E_{cpl}^{tot} = \Delta H_{GNB} - \Delta H_{gr} - \Delta H_{C_{60}}, \qquad (1)$$

where $\Delta H_{GNB}$, $\Delta H_{gr}$, and $\Delta H_{C_{60}}$ determine heats of formation of the considered equilibrium GNB, graphene itself, and fullerene $C_{60}$, respectively. As seen in the table, the GNB formation is accompanied by high coupling energy whose negative sign points to the energetically favorable process.

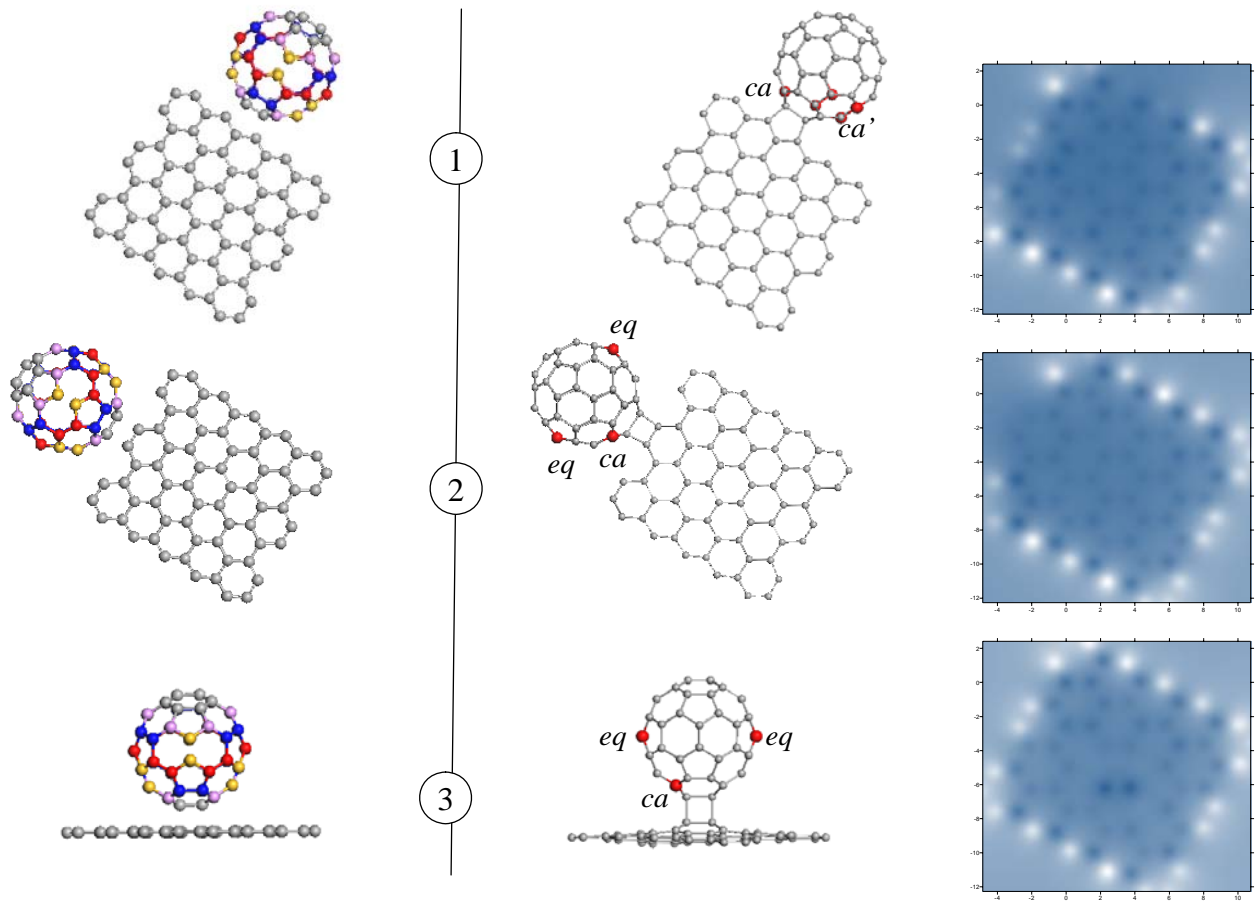

**Figure 2.** [$C_{60}$+(5,5)] graphene nanobuds formed by attaching $C_{60}$ fullerene to zigzag (1) and armchair (2) edge atoms as well as to the basal plane (3) of (5,5) nanographene. Start (left) and equilibrium (right) configurations with ACS maps of the latter. Red balls on equilibrium structures point fullerene atoms with the high-rank ACS values.

**Table 1.** Energetic characteristics of [$C_{60}$+(5,5)] graphene nanobuds, *kcal/mol*

| Composite[1] | $E_{cpl}^{tot}$ | $E_{def}^{tot}$ | $E_{defgr}$ | $E_{defC_{60}}$ | $E_{chem}^{tot}$ |
|---|---|---|---|---|---|
| 1 zg      | -128,45 | 70,56  | 20,7  | 49,86 | -199,01 |
| 2 ach     | -123    | 52,67  | 13,37 | 39,3  | -175,67 |
| 3 b       | -12,11  | 86,8   | 53,76 | 33,04 | -98,91  |
| 4 zgH1    | -52,74  | 109,89 | 69,08 | 40,81 | -162,63 |
| 5 achH1   | -70.17  | 91.53  | 58.10 | 33.44 | -161.71 |
| 6 bH1     | -27,38  | 96,63  | 63,48 | 33,25 | -124,11 |
| 7 bH2     | 1,82    | 62,24  | 28,97 | 33,27 | -60,42  |
| 8[2] bH1  | 32      | 88,98  | 55,75 | 33,23 | -56,98  |

[1] Figures correspond to GNB's numbers in Figs. 2, 4, and 5.
[2] Data for [$C_{60}$+(9,8)] GNB.

    Looking at the ACS map of GNB 1 in Fig.2, one can conclude that the GNB formation causes rather local changes in the map. Two brightly shining zigzag atoms in the left upper corner of the map in Fig.1a are substituted with two dark spots in Fig.2 while ACS of the remainder atoms is practically not altered. This conclusion is justified by the ACS plottings presented in Fig.3. Therefore, from the ACS viewpoint, the response of nanographene to the nanobud formation is quite similar to that one characteristic for carbon nanobuds [3]. Similar as well is the behavior of the attached fullerene that remains still chemically active. Its ACS map considerably changes after addition revealing new target atoms shown by red balls on equilibrium structures. It should be mentioned that the fullerene activity zone is quite similar to that one related to CNB 1 in Fig.6 of Part II [3]. It is quite expected due to zigzag structure of the (4,4) SWCNT open end.

    Graphene nanobuds 2 and 3 in Fig. 2 are formed by attaching fullerene $C_{60}$ either to armchair edge of the graphene sheet or to its basal plane. Analysis of the relevant ACS maps in Fig.2 and their plottings in Fig.3 evidences a local perturbation in the odd electronic state of the graphene caused by the fullerene attachment in both cases. Similarly, both intermolecular junctions are presented by [2+2] cycloadditions. It should be noted therewith that a typical [2+2] junction, identical to that one for $C_{60}$ oligomers [2] and carbon nanobuds [3], takes place in GNB 3 only since in the case of GNB 2 the four-atom contact zone is supported by carbon atoms with two neighbors from the graphene side while the support atoms of fullerene have three neighbors. In spite of the difference, the junctions provide a high similarity in the construction of the fullerene active zones in the two cases. The latter consists of pairs of contact adjacent *ac* and equatorial *eq* atoms and is fully identical to those discussed earlier for [2+2] cycloadducts related to both $C_{60}$ oligomers and CNBs [2, 3]. Any further chemical modification of the GNBs via fullerene will depend on the addend size and will be favorable by targeting *ca* atoms by small addends while becoming preferable when targeting *eq* atoms by bulky addends. In contrast to structural similarity, energetic parameters of the two GNBs are quite different. If those for GNB 2 are similar by value to GNB 1, the coupling energy $E_{cpl}^{tot}$ of GNB 3 is more than 10 times less by absolute value of the previous two.

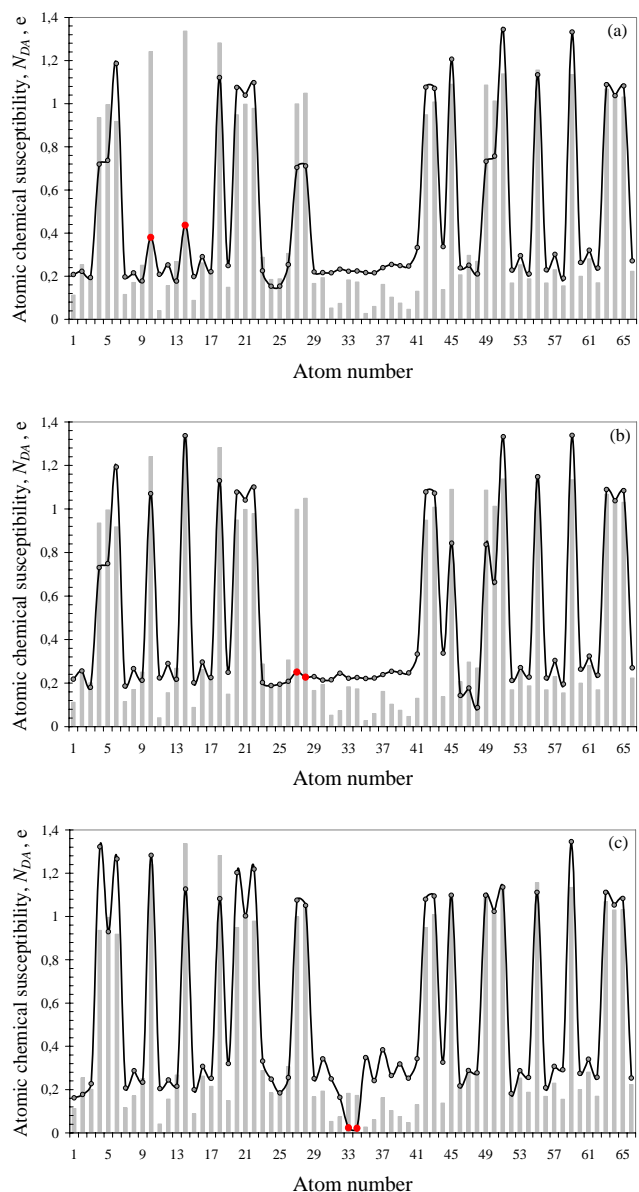

**Figure 3.** Atomic chemical susceptibility distribution over graphene atoms (curves with dots) in [$C_{60}$+(5,5)] graphene nanobuds: GNB 1 (a), GNB 2 (b), and GNB 3 (c). Red balls mark graphene atoms to which $C_{60}$ is covalently attached. Light gray histograms present ACS distribution for the pristine empty-edge (5,5) nanographene.

Local character of the graphene ACS map perturbation provides the formation of not only mono [$C_{60}$+(5,5)] GNBs considered above, but multiple [$(C_{60})_n$+(5,5)] GNBs looking like one of numerous possible examples presented in Fig. 4a. In its turn, attached fullerenes may serve as centers for further chemical modification of the GBDs by the formation of branched chains of different configurations (see Fig. 4b). The variety of possibilities to form different nanobuds based on the same addend is evidently, not a prerogative of fullerene. The same situation will occur with any other adsorbate, once covalently bound with graphene substrate. It is due to a sharp variety of the atomic chemical susceptibility over the pristine graphene sheet, on one hand, and the local character of the graphene sheet perturbation caused by each addition, on the other. This feature would obviously influence, say, the local electric response that will vary depending on place of contact of the same adsorbate thus putting a serious problem concerning reliability of graphene application as selective sensors of different adsorbates [10].

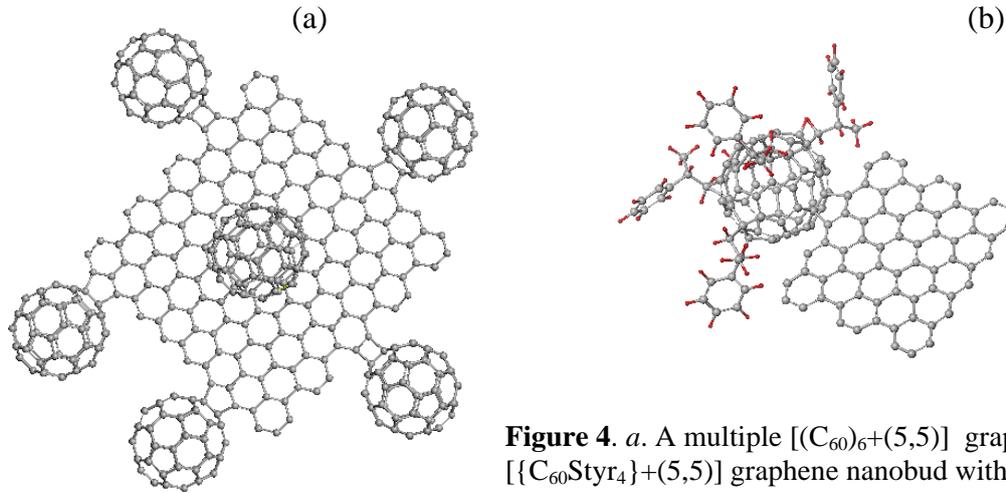

**Figure 4**. *a*. A multiple [$(C_{60})_6$+(5,5)] graphene nanobud. *b*. [{$C_{60}Styr_4$}+(5,5)] graphene nanobud with $C_{60}$-tetrastyrene.

## 2.2. Deposition of $C_{60}$ on nanographene with hydrogen-terminated edges

Figure 5 presents GNBs that can be formed in this case. As seen in the figure, the [$C_{60}$+(5,5)] GNBs behavior is similar to that described in the previous section concerning the local character of the graphene perturbation, the dependence on the place of contact, and the contact zone configurations. In the case of double-H terminated graphene, the activity of zigzag and armchair edges is fully suppressed so that only contacts on the basal plane take place in the GNBs formation. As previously, the contact zones of GNBs 4 and 5 are not explicitly [2+2] cycloaddition junctions while those of GNBs 6 and 7 belong to the latter.

Changes in the chemical activity of the edge atoms greatly influence energetic parameters of the covalent bonding as seen in Table 1. The coupling energies $E_{cpl}^{tot}$ related to the addition to zigzag and armchair edges of single-H terminated graphene decrease more than twice by value. At the same time, addition to the basal plane is accompanied by more than twice increase in the value. The double termination of the graphene edge atoms deprives them of noticeable chemical reactivity leaving the only chance for the covalent bonding with carbon atoms located on basal plane. As seen in Fig. 5, such bonding is formed indeed but needs for its completion ~ 2 kcal/mol.

Graphene nanobud 8 formed by $C_{60}$ covalently coupled with atoms on the basal plane of single-H terminated (9,8) nanographene shown in Fig.6 completes the list of studied GNBs. A typical [2+2] cycloaddition forms the contact zone. This composition presents in our opinion the situation that should be often met in real practice. Nanographenes of *nm* in size with definitely terminated edge atoms might be expected as the most abundant part of the graphene solutions [11]. Nanobuds formed at the edges of the sheet behave quite similarly those formed on (5,5) sheet discussed above. However, the formation of GNB 8 highlights a new feature since the coupling energy is high by value but positive, which means that the GNB formation is not energetically profitable in contrast to GNB 6. To elucidate possible reasons for such behavior let us examine energetic characteristics of GNB 8 in details.

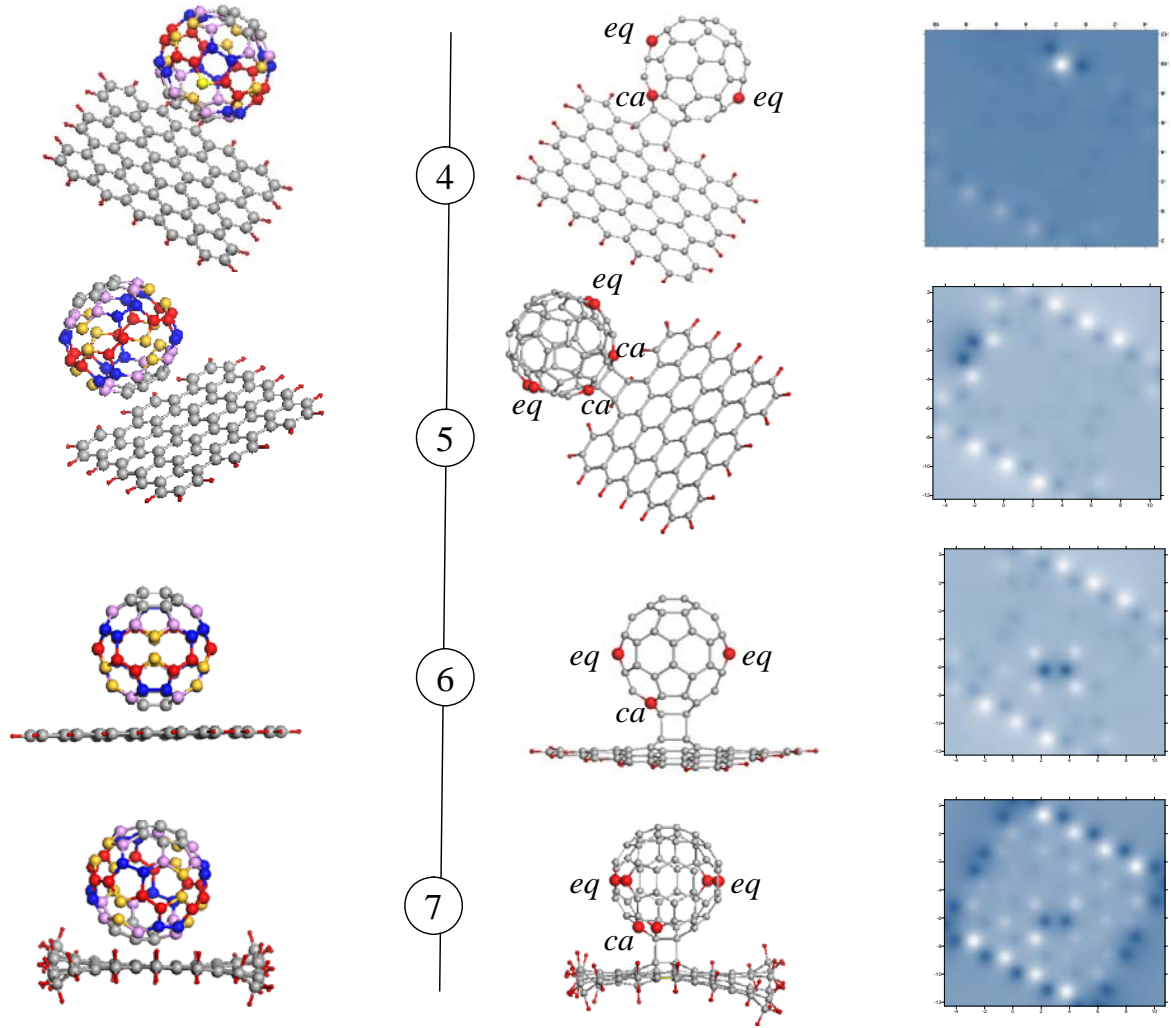

**Figure 5.** [$C_{60}$+(5,5)] graphene nanobuds formed by attaching $C_{60}$ fullerene to zigzag (4) and armchair (5) edge atoms as well as to the basal plane (6) and (7) in the case of single-H (4-6) and double-H (7) terminated (5,5) nanographene. Start (left) and equilibrium (right) configurations with ACS maps of the latter. Red balls on equilibrium structures point to fullerene atoms with the high-rank ACS values.

## 3. Energetic parameters and reaction barrier of graphene nanobuds

As was discussed in Parts I, and II, it is quite reasonable to present the total coupling energy $E_{cpl}^{tot}$ related to GNBs consisting of two components, namely, $E_{def}^{tot}$ and $E_{chem}^{tot}$ that take the form

$$E_{def}^{tot} = E_{defgr} + E_{defC_{60}}, \qquad (2)$$

where

$$E_{defgr} = \Delta H_{gr}^{GNB} - \Delta H_{gr} \quad \text{and} \quad E_{defC_{60}} = \Delta H_{C_{60}}^{GNB} - \Delta H_{C_{60}}. \tag{3}$$

Here $\Delta H_{gr}^{GNB}$ and $\Delta H_{C_{60}}^{GNB}$ present heats of formation of one-point-geometry configurations of the graphene and fullerene components of the equilibrium configurations of the studied GNB. Accordingly, the chemical contribution into the coupling energy can be determined following the relation

$$E_{\text{cov}}^{tot} = E_{cpl}^{tot} - E_{def}^{tot}. \tag{4}$$

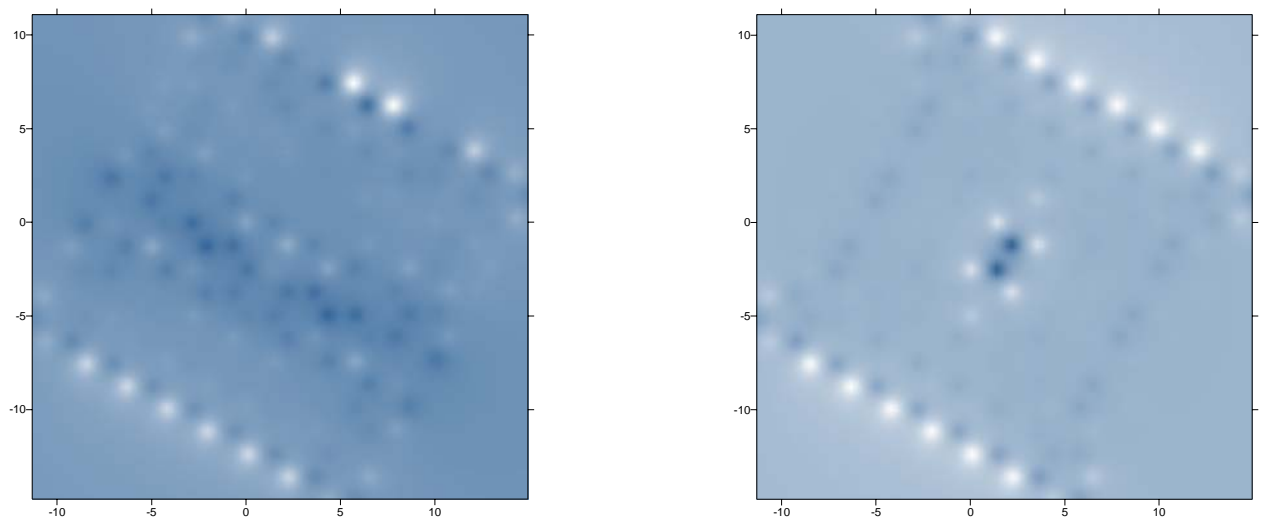

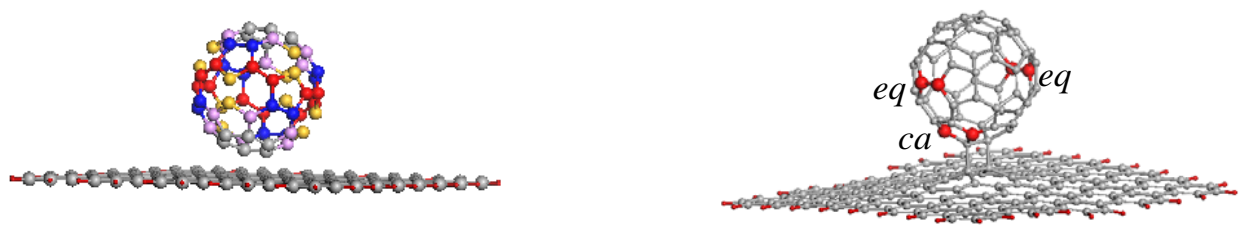

**Figure 6.** [$C_{60}$+(9,8)] graphene nanobud formed by attaching $C_{60}$ fullerene to the basal plane of (9,8) nanographene. Start (left) and equilibrium (right) configurations with ACS maps (above) related to initial and perturbed graphene. Red balls on equilibrium structures point to fullerene atoms with the high-rank ACS values.

Figure 7 presents the dependence of $E_{cpl}^{tot}$, $E_{def}^{tot}$ and $E_{chem}^{tot}$ on the intermolecular 1-1` and 2-2` C-C distances (see Fig.2a in Part I) for GNB 8. The three plottings in the figure are generally similar to those presented in Fig. 3 of Part I and Fig. 8 of Part II related to $C_{60}$ dimer and carbon nanobuds, respectively. This is obviously resulted from the similarity of atomic structure of the contact zones formed in all the considered cases by [2+2] cycloadditions. The coupling energy $E_{cpl}^{tot}$ can be evidently divided into $E_{def}^{tot}$ and $E_{cov}^{tot}$ components of the same type as those related to the previous two cases. However, the difference in numerical values of the two components at start point results in a remarkable lifting of the $E_{cpl}^{tot}$ minimum for GNB 8 moving it into positive energy region. Consequently the barrier energy $E_{barr}^{GNB}$ lowers up to 22.7 kcal/mol. Analyzing the data, one can conclude that the feature is mainly caused by a drastic lowering of the covalent contribution into $E_{cpl}^{tot}$ in this case.

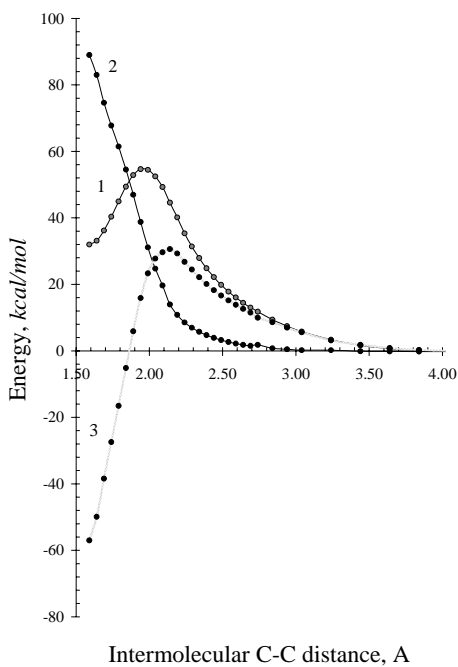

**Figure 7.** Profile of the barrier of the [$C_{60}$+(9,8)] GNB decomposition. 1. $E_{cpl}^{tot}$; 2. $E_{def}^{tot}$; 3. $E_{cov}^{tot}$.

## 4. Discussion

The studied nanobuds, including $C_{60}$ oligomers, carbon nanobuds, and GNBs, are resulted from a covalent pair-pair bonding between the components, where each member of the latter delegates a pair of the most chemically active atoms to form intermolecular junctions. The relevant atom pairs have been selected following the indication of the high-rank ACS atoms on the corresponding ACS maps. A high non-homogeneity in the ACS distribution over atoms of SWNT and graphene, which divides their space into three regions, namely: cap, end, and sidewall of CNT as well as zigzag and armchair edges and basal plane of graphene, forms the grounds for the dependence of the formed nanobuds from the $C_{60}$ place of location and from the state-of-art termination of both end and edge atoms. Consequently, the formed nanobuds present a rather complicated set of covalently bound composites that differ by both coupling energies and the structure of intermolecular junctions. Thus, only $C_{60}$ dimer, sidewall carbon nanobuds, and basal-plane GNBs can be characterized by the [2+2] cycloaddition as alike intermolecular junction. Table 2 summarizes data covering the three studied cases.

As seen in Table 2, structural characteristics concerning the [2+2] cycloadditions are practically identical within carbon nanobuds and GNB groups while the coupling energies differ therewith quite considerably. It is difficult to connect the difference with rather small variance concerning $N_{DA}$ values. Therefore, in contrast to a close similarity in both coupling and barrier energies for the studied nanobuds that might be expected in the cases, we see a large variety of properties. It is hidden from view now what is the reason for the observed peculiarity. Obviously, it is not connected with the computational procedure since in all the cases when the obtained data could be compared with experimental ones (dimers and oligomers [9]) a perfect fitting was obtained. The data generally agree as well with the available computational results [1, 12, 13] so that we are facing some unknown peculiarity in the chemical behavior of the studied $sp^2$ nanocarbons. Obviously, the observed feature is connected with the specificity of intermolecular interaction between the components of the composites. In view of this, two questions can be addressed. 1) Might be is wrong our suggestion concerning the superposition of deformational and chemical contributions into the total coupling energy? However, a standard view of $E_{def}^{tot}(R_{CC})$ and $E_{cov}^{tot}(R_{CC})$ plottings in Fig. 3 of Part I, Fig. 8 of Part II, and Fig.7 appears to support this suggestion. If the suggestion is true, 2) might be a new third contribution is not taken into account, which so much effects $E_{cov}^{tot}$? Actually, as seen in Fig.8 where the three $E_{cov}^{tot}(R_{CC})$ dependences are joined together, a feeling arises that each of them well reproduces the others but is shifted along the energy scale. It seems reasonable to suggest that the shift just exhibits an anonymous contribution. The considered nanocarbons are peculiar species, obviously not studied up to the bottom. Concerning the current case such a peculiarity of the species as their topology has not been so far taken into account. At the same time, one cannot exclude revealing some peculiar components of the intermolecular interaction similar to Casimir's repulsion in topological insulators [14] since the studied nanobuds are evidently formed under different topological conditions.

**Table 2.** Joint characteristics of the nanobud [2+2] cycloadditions

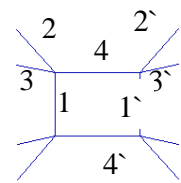

| Nanobuds | C-C bonds, Å[1] | | | | $N_{DA}$[2], e | $E_{chem}^{tot}$, kcal/mol | $E_{barr}^{NB}$, kcal/mol |
|---|---|---|---|---|---|---|---|
| | 1 | 2 | 3 | 4 | | | |
| $C_{60}+C_{60}$[3] | 1.548 | 1.515 | 1.515 | 1.596 | 0.271, 0.271 | -112.97 | 65.24 |
| | 1.548 | 1.516 | 1.516 | 1.596 | | | |
| CBN 3[4] | 1.567 | 1.483 | 1.486 | 1.652 | 0.308, 0.307 | -134,31 | - |
| | 1.567 | 1.520 | 1.518 | 1.590 | | | |
| CBN 5[4] | 1.566 | 1.484 | 1.486 | 1.652 | 0.305, 0.304 | -78,59 | 36.07 |
| | 1.566 | 1.520 | 1.518 | 1.590 | | | |
| GBN 3[4] | 1.591 | 1.496 | 1.496 | 1.581 | 0.183, 0.173 | -98,91 | - |
| | 1.591 | 1.519 | 1.518 | 1.579 | | | |
| GBN 6[4] | 1.589 | 1.493 | 1.494 | 1.578 | 0.216, 0.201 | -124,11 | - |
| | 1.589 | 1.519 | 1.517 | 1.580 | | | |
| GBN 7[4] | 1.590 | 1.492 | 1.492 | 1.578 | 0.189, 0.165 | -60,42 | - |
| | 1.589 | 1.517 | 1.519 | 1.580 | | | |
| GBN 8[5] | 1.591 | 1.494 | 1.494 | 1.576 | 0.227, 0.174 | -56.98 | 22.70 |
| | 1.592 | 1.519 | 1.518 | 1.580 | | | |

[1] The bond numeration corresponds to the insert. Two-row presentation distinguishes the primed bonds (the second rows) related to $C_{60}$ in all cases from unprimed ones (the first rows) related to the $C_{60}$ partner.
[2] The data are related to the pairs of atoms of $C_{60}$ partner's in the nanobuds. The data for fullerene partner are presented in the first row.
[3] $C_{60}$ dimer.
[4] [$C_{60}$+(4,4)] CNB. CNB' numbering corresponds to that in Figs. 4 and 5 (Part 2)
[5] [$C_{60}$+(5,5)] GNB. GNB' numbering corresponds to that in Figs. 2 and 5.
[6] [$C_{60}$+(9,8)] GNB (see Fig.6).

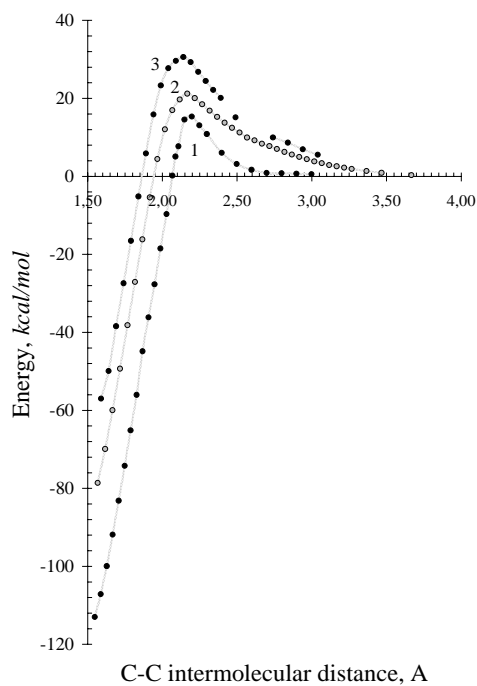

**Figure 8.** $E_{cov}^{tot}$ plotting for $C_{60}$ dimer (1), [$C_{60}$+(4,4)] CNB 5 (2), and [$C_{60}$+(9,8)] GNB 8 (3).